\documentclass[letterpaper,10pt]{article}

\usepackage{amsmath}
\usepackage{amssymb}
\usepackage{psfrag}
\usepackage[dvips]{graphicx}
\usepackage{epsfig}

\newcommand{\dirac}{{\slash \negthinspace \negthinspace \negthinspace \nabla}}
\newcommand{\dd}{\textrm{d}}
\newcommand{\im}{{\mathbb{I}}{\mathrm{m}}}
\newcommand{\re}{{\mathbb{R}}{\mathrm{e}}}

\author{R.\ Cordero\thanks{cordero@esfm.ipn.mx}   \\
Departamento de F\'{\i}sica. Escuela Superior de F\'{\i}sica y Matem\'aticas. \\
Instituto Polit\'ecnico Nacional. \\
Unidad Profesional Adolfo L\'opez Mateos, Edificio 9. \\
M\'exico, D.\ F., M\'exico. \\
C.\ P.\ 07738
\and
A. L\'opez-Ortega\thanks{alopezo@ipn.mx} \\
Centro de Investigaci\'on en Ciencia Aplicada y Tecnolog\'{\i}a Avanzada. \\ 
Unidad Legaria. Instituto Polit\'ecnico Nacional. \\
Calzada Legaria \# 694. Colonia Irrigaci\'on. Delegaci\'on Miguel Hidalgo. \\
M\'exico, D.\ F., M\'exico. \\
C.\ P.\  11500  
\and
I.\ Vega-Acevedo\thanks{ivega@esfm.ipn.mx}\\
Departamento de F\'{\i}sica. Escuela Superior de F\'{\i}sica y Matem\'aticas. \\
Instituto Polit\'ecnico Nacional. \\
Unidad Profesional Adolfo L\'opez Mateos, Edificio 9. \\
M\'exico, D.\ F., M\'exico. \\
C.\ P.\ 07738
}

\title{Quasinormal frequencies of asymptotically anti-de Sitter black holes in two dimensions}

\begin{document}

\maketitle

\begin{abstract}

We calculate exactly the quasinormal frequencies of Klein-Gordon and Dirac test fields propagating in two-dimensional uncharged Achucarro-Ortiz black hole. For both test fields we study whether the quasinormal frequencies are well defined in the massless limit. We use their values to discuss the classical stability of the quasinormal modes in uncharged Achucarro-Ortiz black hole and to check the recently proposed Time Times Temperature bound. Furthermore we extend some of these results to the charged Achucarro-Ortiz black hole.

\vspace{.3cm}



\end{abstract}

\section{Introduction}
\label{s: Introduction}

In classical physics, the black holes are intrinsically dissipative systems because the one way property of the event horizon implies that the perturbations travel only to the black hole interior. Depending on the asymptotic structure of the spacetime the perturbations can escape to infinity or satisfy other boundary conditions at the asymptotic region.

The modes of the perturbations that satisfy the natural radiation boundary conditions at the horizon and the infinity are called quasinormal modes (QNM). Their associated frequencies are complex and are usually called  quasinormal frequencies (QNF). It is well known that the QNF are determined by the parameters of the black hole as the mass, charge, and angular momentum. For more details see the extensive reviews \cite{Kokkotas:1999bd}--\cite{Konoplya:2011qq}.

Therefore the QNM give us an useful tool for determining the black hole parameters. In addition to the QNF of black holes that are relevant for astrophysics, the QNF of other gravitational backgrounds have been calculated. In particular, motivated by the AdS-CFT correspondence, the QNF of asymptotically anti-de Sitter black holes are thoroughly studied (see references \cite{Horowitz:1999jd}--\cite{Sachs:2008gt} for some examples and the reviews \cite{Berti:2009kk,Konoplya:2011qq} for additional references), since according to the AdS-CFT correspondence the QNF of asymptotically anti-de Sitter black holes are related to the hydrodynamic properties of a gauge theory fluid.

In many references are used perturbative or numerical methods to calculate the QNF of the black holes  (see the reviews \cite{Kokkotas:1999bd}--\cite{Konoplya:2011qq} and references therein). Nevertheless for some spacetimes their QNF are computed exactly and it should be noted that many of these gravitational systems are lower dimensional. In three dimensions we know the BTZ black hole \cite{Birmingham:2001pj}--\cite{Sachs:2008gt}, the warped AdS(3) black hole \cite{Chen:2009hg}--\cite{Li:2010sv}, the dilatonic black hole \cite{Fernando:2003ai}--\cite{Fernando:2009tv}, and the de Sitter spacetime \cite{Du:2004jt,Lopez-Ortega:2006ig}.

For two-dimensional black holes we know the exact calculations of the QNF that appear in Refs.\ \cite{Zelnikov:2008rg}--\cite{LopezOrtega:2011sc}. Zelnikov \cite{Zelnikov:2008rg} calculates exactly the QNF of the Klein-Gordon field propagating in a family of two-dimensional black holes. Becar et al.\  \cite{Becar:2007hu} compute exactly the QNF of the Klein-Gordon field minimally coupled to scalar curvature in the uncharged Witten black hole  \cite{Witten:1991yr,Mandal:1991tz} (see also \cite{LopezOrtega:2009zx}). For this black hole, also Becar et al.\ \cite{Becar:2010zz} calculate exactly the QNF of the massless Dirac field and in \cite{LopezOrtega:2011sc} these results are extended to the massive Dirac field. Furthermore in the last reference it is shown that in asymptotically flat two-dimensional black holes the massless Klein-Gordon and Dirac fields do not have well defined QNM and hence the frequencies calculated in \cite{Becar:2010zz} cannot be QNF of the massless Dirac field in the uncharged Witten black hole.\footnote{See also Refs.\ \cite{Kettner:2004aw,Li:2001ct} to find the analytical calculation of the asymptotic QNF for the coupled to dilaton Klein-Gordon field in one family of two-dimensional black holes \cite{Kettner:2004aw} and the computation with the WKB method of the first QNF for the charged Witten black hole \cite{Li:2001ct}. }

Taking into account the known advantages of the two-dimensional gravity models in the analysis of many physical phenomena  \cite{Grumiller:2002nm,Grumiller:2006rc}, it is unexpected that for two-dimensional black holes their QNF have not been studied in more detail. Motivated by this fact here we extend the previous results and calculate the QNF of the uncharged Achucarro-Ortiz\footnote{Sometimes this spacetime is called AdS$_2$ black hole \cite{Kim:1998wy}--\cite{Kim:1999ig}.} (UAO) black hole and of the charged Achucarro-Ortiz (CAO) black hole \cite{Achucarro:1993fd}. In contrast to the black holes studied in Refs.\ \cite{Zelnikov:2008rg}--\cite{LopezOrtega:2011sc}, \cite{Kettner:2004aw,Li:2001ct}  the UAO and CAO black holes are asymptotically anti-de Sitter. Therefore the boundary condition that the fields satisfy at the asymptotic region is different from the boundary condition imposed in \cite{Zelnikov:2008rg}--\cite{LopezOrtega:2011sc}, \cite{Kettner:2004aw,Li:2001ct}. Although for the two black holes in two dimensions that we study here, their metrics are similar to those of the three-dimensional static and rotating BTZ black holes, in the results for their QNF we find some differences respect to those naively expected.

We organize this paper as follows. In Section \ref{s: AdS(2) black hole} we calculate exactly the QNF of the Klein-Gordon and Dirac fields propagating in UAO black hole. Using these results we test the classical stability of the QNM and the Time Times Temperature bound. In Section \ref{s: Charged AO black hole} we extend numerically some of the previous results to the CAO black hole. Finally in Section \ref{s: Summary} we discuss the obtained results.

\section{Quasinormal frequencies of the UAO black hole}
\label{s: AdS(2) black hole}

Among the two-dimensional black holes, the UAO black hole is frequently studied \cite{Grumiller:2002nm,Grumiller:2006rc,Achucarro:1993fd}. The metric and dilaton of this black hole are
\begin{equation} \label{e: metric AdS(2) black hole}
\dd s^{2} =\left(-M + \frac{r^{2}}{l^{2}}\right) \dd t^{2}- \left(-M + \frac{r^{2}}{l^{2}}\right)^{-1} \dd r^{2}, \qquad \qquad \phi = r,
\end{equation}
where $M$ is the mass of the black hole and $l^2$ is related to the three-dimensional cosmological constant (see below). Taking $r_+^2 = M l^2$  and replacing $t/l^2$ by $t$ we write the metric (\ref{e: metric AdS(2) black hole}) in the form
\begin{equation} \label{e: metric AdS(2) conformal}
 \dd s^2 = l^2 \left( (r^2 -r_+^2) \dd t^2 - \frac{\dd r^2}{(r^2 -r_+^2) } \right),
\end{equation}
and instead the line element (\ref{e: metric AdS(2) black hole}) we take
\begin{equation} \label{e: metric AdS(2)}
 \dd s^2 =  (r^2 -r_+^2) \dd t^2 - \frac{\dd r^2}{(r^2 -r_+^2) } ,
\end{equation}
as the metric of the UAO black hole, because the spacetimes (\ref{e: metric AdS(2) black hole}) and (\ref{e: metric AdS(2)}) are related by a simple conformal transformation. Note that if we take $l=1$ and hence $r_+^2=M$, the metrics (\ref{e: metric AdS(2) black hole}) and (\ref{e: metric AdS(2)}) are equal.

We can get this two-dimensional spacetime from the three-dimensional static BTZ black hole \cite{Banados:1992wn}--\cite{Carlip:1995qv} by means of a dimensional reduction procedure, in such a way that the UAO black hole is a solution to the equations of motion derived from the action \cite{Achucarro:1993fd}
\begin{equation}
 S = \int \dd^2 x \sqrt{|g|} \,\, \phi \left( \mathcal{R} + \frac{2}{l^2} \right),
\end{equation}
where $\mathcal{R}$ is the scalar curvature and $\Lambda= 1/ l^2$ is the three-dimensional cosmological constant.

The propagation of fields in UAO spacetime is studied in Refs.\ \cite{Kim:1998wy}--\cite{Kim:1999ig}, \cite{Lifschytz:1993eb}. For test fields moving in UAO black hole we define their QNM as the oscillations that satisfy:
\begin{description}
\item[a)] The oscillations are purely ingoing near the event horizon.

\item[b)] The oscillations go to zero as $r \to \infty$.
\end{description}
The boundary conditions a) and b) are similar to those for the QNM of asymptotically anti-de Sitter black holes \cite{Berti:2009kk,Konoplya:2011qq} as the three-dimensional BTZ black hole \cite{Birmingham:2001pj,Cardoso:2001hn,Birmingham:2001hc} and the Schwarzschild anti-de Sitter black hole \cite{Horowitz:1999jd}--\cite{Chan:1996yk}. In what follows we calculate exactly the QNF of Klein-Gordon and Dirac test fields that propagate in UAO black hole.

\subsection{Quasinormal frequencies of the Klein-Gordon field}
\label{s: QNF Klein-Gordon}

Let us calculate the QNF of the Klein-Gordon field in UAO black hole. As it is well known the massive scalar field satisfies the Klein-Gordon equation
\begin{equation} \label{e: Klein Gordon}
\left( \square + m^{2} \right) \Phi = 0,
\end{equation}
where $\square$ is the two-dimensional d'Alembertian and $m$ denotes the mass of the field. We notice that for the UAO black hole the scalar curvature is a constant, hence it is straightforward to consider a coupling of the Klein-Gordon field to scalar curvature since we  need only to  define an effective mass determined by the mass of the field and the constant scalar curvature times the coupling constant.

Proposing that the Klein-Gordon field takes the form
\begin{equation}
\label{e: separation KG}
\Phi(r,t) = R(r) e^{-i\omega t},
\end{equation}
in the metric (\ref{e: metric AdS(2)}) the Klein-Gordon equation reduces to the radial differential equation
\begin{equation} \label{e: radial Klein Gordon}
z(1-z) \frac{\dd^{2}R}{\dd z^{2}} + \left( 1 - \frac{3 z}{2} \right)\frac{\dd R }{\dd z} + \left(\frac{\alpha}{z} + \frac{\beta}{1-z}\right) R = 0,
\end{equation}
with
\begin{align} \label{e: alpha beta Klein-Gordon}
z = \frac{r^{2}-r^{2}_{+}}{r^{2}} ,\qquad \qquad
\alpha = \frac{\omega^{2} }{ 4 r^2_+ } ,\qquad \qquad
\beta = -\frac{m^2}{4} .
\end{align}

For the radial function $R$ we make the ansatz
\begin{equation} \label{e: radial function ansatz}
R(z) = z^{C} (1-z)^{B} Q(z),
\end{equation}
where
\begin{equation} \label{e: chosen values B C}
 C =- \frac{i\omega}{2 r_+} , \qquad \qquad  B  = \frac{1}{4} - \frac{1}{4} \sqrt{1 + 4 m^{2} } ,
\end{equation}
to find that the function $Q$ is a solution of the hypergeometric differential equation \cite{Abramowitz-book,Guo-book}
 \begin{equation} \label{e: hypergeometric equation}
 z(1-z)\frac{\dd^2 Q}{\dd z^2} + (c -(a+b+1)z)\frac{\dd Q}{\dd z} - a b Q = 0 ,
\end{equation}
with parameters $a$, $b$, and $c$ equal to
\begin{align} \label{e: parameters hypergeometric}
a  =  B + C + \tfrac{1}{2} , \qquad \quad
b  = B + C , \qquad \quad
c  =  2 C + 1.
\end{align}

If the quantity $c$ is not an integer and taking into account the time dependence of the Klein-Gordon field (\ref{e: separation KG}) we get that near the horizon the radial function that satisfies the boundary condition of the QNM  is
\begin{equation} \label{e: radial AdS(2) horizon}
R = E \, z^{- i \omega / 2 r_+} (1-z)^{B}  {}_{2}F_{1}(a,b;c;z),
\end{equation}
where $E$ is a constant and ${}_{2}F_{1}(a,b;c;z)$ denotes the hypergeometric function \cite{Abramowitz-book,Guo-book}.

We point out that $z \to 1$ as $r \to \infty$ and if $m \neq 0$ then $B < 0$ for the value chosen in the expressions (\ref{e: chosen values B C}), thus in the radial function (\ref{e: radial AdS(2) horizon}) the term $(1-z)^B$ diverges as $z \to 1$. Furthermore we note that for the hypergeometric function in the formula (\ref{e: radial AdS(2) horizon}) its parameters $a$, $b$, and $c$  satisfy $c- a- b = \sqrt{1 + 4 m^2} / 2 > 0$. Therefore we can use that for $\re (c -a - b) > 0$ the hypergeometric function fulfills \cite{Guo-book}
\begin{equation} \label{e: value hypergeometric 1}
 \lim_{z \to 1}  {}_{2}F_{1}(a,b;c;z) = \frac{\Gamma(c) \Gamma(c-a-b)}{\Gamma(c-a) \Gamma(c-b)}.
\end{equation}

Thus to satisfy the boundary condition at infinity we must solve the equations
\begin{equation} \label{e: conditions QNF}
c-a=-n, \qquad \textrm{or} \qquad c-b=-n, \qquad n=0,1,2,\dots
\end{equation}
From these conditions we get that in UAO black hole the QNF of the massive Klein-Gordon field are equal to
\begin{equation}\label{e: QNF AdS(2) KG}
 \omega = -i r_+ \left(n + \tfrac{1}{2} + \tfrac{1}{2}\sqrt{1 + 4 m^{2}} \right) .
\end{equation}
Note that the QNF of the Klein-Gordon field in UAO black hole are purely imaginary as those of Refs.\ \cite{Fernando:2003ai}--\cite{LopezOrtega:2011sc} for lower dimensional gravitational systems.

To compare the QNF of the two-dimensional UAO black hole and the three-dimensional static BTZ black hole, we point out that in the latter the QNF of the Klein-Gordon field are\footnote{For the static BTZ black hole we take $l=1$, and as previously, $m$ denotes the mass of the Klein-Gordon field; in Ref.\ \cite{Birmingham:2001hc} the mass of the Klein-Gordon field is denoted by $\sqrt{\mu}/l$. }
\begin{equation} \label{e: QNF BTZ Klein Gordon}
 \omega_{BTZ} = \pm k - i r_+ \left( 2 n + 1 + \sqrt{1+m^2} \right),
\end{equation}
(see the formula (18) of \cite{Birmingham:2001hc}) and notice that $k$ denotes the azimuthal number in the ansatz used to make the separation of variables in static BTZ black hole \cite{Birmingham:2001pj,Cardoso:2001hn}.

Notice that the factor $4 m^2$ in the square root of the QNF (\ref{e: QNF AdS(2) KG}) is different from the similar factor $m^2$ that appear in the square root of the QNF (\ref{e: QNF BTZ Klein Gordon}) for the static BTZ black hole. Also notice that in the QNF (\ref{e: QNF AdS(2) KG}) the term $n + 1/2$ is a half-integer, whereas in the QNF (\ref{e: QNF BTZ Klein Gordon}) of the static BTZ black hole the similar term is an integer. Hence the QNF of the massive Klein-Gordon field in UAO black hole cannot be obtained as the limit $k \to 0$ from those of the three-dimensional static BTZ black hole.

From the formula (\ref{e: separation KG}) we get that the Klein-Gordon field has a harmonic time dependence of the form $\exp(-i \omega t)$, thus to have fields that decay in time we need that the imaginary parts of the QNF fulfill $\im \,\omega < 0$. For the QNF of the Klein-Gordon field in UAO black hole, since $ n + \tfrac{1}{2} + \tfrac{1}{2} \sqrt{1+4m^2} > 0 ,$ we observe that their imaginary parts satisfy $\im\, \omega < 0$ and therefore in UAO black hole the QNM of the Klein-Gordon field are classically stable.

In Ref.\ \cite{LopezOrtega:2011sc} it is shown that the massless Klein-Gordon field does not have well defined QNF in single horizon asymptotically flat two-dimensional black holes. In this reference it is shown that the solutions to the equation of motion do not satisfy the boundary conditions of the QNM for asymptotically flat black holes.

In a similar way to \cite{LopezOrtega:2011sc}, we see that for asymptotically anti-de Sitter two-dimensional black holes the massless Klein-Gordon equation simplifies to a free Schr\"odinger type equation (see Eqs.\ (25) and (26) in \cite{LopezOrtega:2011sc}), and since its solutions do not satisfy the boundary conditions of the QNM, we find that the massless Klein-Gordon field does not have well defined QNF in asymptotically anti-de Sitter two-dimensional black holes.

Another method to show this fact for the UAO black hole is to note that in the massless limit the solutions of the radial equation (\ref{e: radial Klein Gordon}) can be expanded in terms of Jacobi polynomials \cite{Szego-book}. Using these solutions it is possible to demonstrate that we cannot satisfy the boundary conditions of the QNM at the horizon and at infinity.

The previous comments imply that  in the limit $m \to 0$ of the QNF (\ref{e: QNF AdS(2) KG}) we get frequencies that are not quasinormal, because we cannot fulfill the boundary conditions of the QNM for the UAO black hole. Thus the massless Klein-Gordon field does not have well defined QNM in this black hole.

\subsection{Quasinormal frequencies of the Dirac field}
\label{s: QNF Dirac}

In this Subsection we calculate exactly the QNF of the Dirac field in UAO black hole and discuss similar issues to those of the Klein-Gordon field that we analyze in the previous Subsection.  The Dirac equation is
\begin{equation}
 i \dirac \Psi = m \Psi ,
\end{equation}
where $\Psi$ is a two-spinor, $\dirac$ denotes the Dirac operator, and $m$ is the mass of the fermion field. The Dirac operator is given by $\dirac = \gamma^\mu \nabla_\mu$ with $\nabla_\mu$ denoting the covariant derivative and the matrices $\gamma^\mu$ satisfy $\gamma^\mu \gamma^\nu + \gamma^\nu \gamma^\mu = 2 g^{\mu \nu}$.

We notice that for a two-dimensional spacetime with metric
\begin{equation} \label{e: metric general}
\dd s^{2}=f\dd t^{2}-\frac{\dd r^{2}}{g},
\end{equation}
the Dirac equation in the chiral representation  simplifies to \cite{LopezOrtega:2011sc}
\begin{align} \label{e: Dirac coupled}
\partial_{t} \Psi_{2} - \partial_{r_*} \Psi_{2} = -i m \sqrt{f} \Psi_{1}, \qquad \qquad
\partial_{t} \Psi_{1}  + \partial_{r_*} \Psi_{1} = -i m \sqrt{f} \Psi_{2} ,
\end{align}
where the functions $\Psi_1$ and $\Psi_2$ are related to the components of the two-spinor $\Psi$, and $r_*$ is the tortoise coordinate.

In UAO black hole we find that Eqs.\ (\ref{e: Dirac coupled}) transform into
\begin{align} \label{e: radial Dirac 1 2}
 \frac{\dd^2 P_{1|2}}{\dd z^2} &+ \left( \frac{1}{2z} - \frac{1}{1-z} \right) \frac{\dd P_{1|2}}{\dd z}   \\
&+ \left( \frac{(\tilde{\omega}^2 \pm i \tilde{\omega} ) - (m + 1/2)^2}{4 z (1-z)}
+ \frac{ \tilde{\omega}^2 \pm i \tilde{\omega} }{4 z^2} - \frac{(m + 1/2)^2}{4 (1-z)^2} \right) P_{1|2} = 0, \nonumber
\end{align}
where $\Psi_{1|2}= R_{1|2}e^{-i\omega t}$, $\tilde{R}_1  = - i R_1 $, $\tilde{\omega} = \omega / r_+$,
\begin{eqnarray}
 R_2 + \tilde{R}_1 = (1 - \tanh^2(\sigma) )^{-1/4} (1 + \tanh(\sigma))^{1/2} (P_2 + P_1), \nonumber \\
R_2 - \tilde{R}_1 = (1 - \tanh^2(\sigma) )^{-1/4} (1 - \tanh(\sigma))^{1/2} (P_2 - P_1),
\end{eqnarray}
$r = r_+ \cosh (\sigma) $, $y = \tanh (\sigma)$, and $z = y^2$ \cite{Dasgupta:1998jg}. We observe that in Eq.\ (\ref{e: radial Dirac 1 2}) the upper signs refer to the function $P_1$ and the lower signs to the function $P_2$.

In a similar way to the Klein-Gordon equation we propose that the functions $P_1$ and $P_2$ take the form
\begin{equation} \label{e: ansatz P Dirac}
 P_{1|2} = z^{C_{1|2}} (1-z)^{B_{1|2}} S_{{1|2}} ,
\end{equation}
to find that the functions $S_{{1|2}}$ must be solutions of the hypergeometric type differential equations (\ref{e: hypergeometric equation}) with parameters
\begin{align} \label{e: parameters hypergeometric Dirac}
a_{1|2}  =  B_{1|2} + C_{1|2} + \tfrac{1}{2} , \qquad \quad
b_{1|2}  = B_{1|2} + C_{1|2} , \qquad \quad
c_{1|2}  =  2 C_{1|2} + \tfrac{1}{2}.
\end{align}

We study in detail the component $\Psi_1$ of the Dirac field and then comment on the results for the component $\Psi_2$. Choosing
\begin{equation}
 C_1 = \frac{1}{2} - \frac{i \tilde{\omega}}{2}, \qquad \qquad B_1 = - \frac{m}{2} -\frac{1}{4},
\end{equation}
we find that the radial function that satisfies the boundary conditions at the horizon is
\begin{equation} \label{e: radial Dirac horizon}
P_1 = E_1 z^{(1- i \tilde{\omega} ) / 2 } (1-z)^{- (m/2+1/4)}  {}_{2}F_{1}(a_1,b_1;c_1;z).
\end{equation}

Since the parameters of the hypergeometric function in Eq.\ (\ref{e: radial Dirac horizon}) fulfill
\begin{equation}
{ \mathbb R e} ( c_1-a_1-b_1 )  = \tfrac{1}{2} + m > 0 ,
\end{equation}
in a similar way to the massive Klein-Gordon field we impose the boundary condition at infinity to get that the QNF for the component $\Psi_1$ of the Dirac field in UAO black hole are
\begin{equation} \label{e: QNF Dirac 1 AdS(2)}
 \omega_1 = -i r_+ (m + \tfrac{3}{2} + n).
\end{equation}
A similar method gives that for the component $\Psi_2$ of the Dirac field its QNF are equal to\footnote{To compute the QNF (\ref{e: QNF Dirac 2 AdS(2)}) we take $C_2 = -i \tilde{\omega}/2$ and $B_2 = -m/2 -1/4$.}
\begin{equation} \label{e: QNF Dirac 2 AdS(2)}
 \omega_2 = -i r_+ (m + \tfrac{1}{2} + n) .
\end{equation}

Except for the fundamental mode of $\omega_2 $, the QNF (\ref{e: QNF Dirac 1 AdS(2)}) of the component $\Psi_1$  and the QNF (\ref{e: QNF Dirac 2 AdS(2)}) of the component $\Psi_2$  yield the same values. It is expected since in two-dimensional metrics of the form (\ref{e: metric general}) the Dirac equation simplifies to two Schr\"odinger type equations with potentials that are SUSY partners \cite{Cooper:1994eh} (see the formulas (31) and (32) in \cite{LopezOrtega:2011sc}).

We note that the QNF (\ref{e: QNF Dirac 1 AdS(2)}) and (\ref{e: QNF Dirac 2 AdS(2)}) of the Dirac field are purely imaginary, but different from the QNF (\ref{e: QNF AdS(2) KG}) for the massive Klein-Gordon field. We point out that for the three dimensional static BTZ black hole the QNF of the Dirac field are equal to \cite{Birmingham:2001pj,Cardoso:2001hn}
\begin{equation} \label{e: QNF BTZ Dirac}
 \omega_{BTZ} = k - i r_+ (2n + \tfrac{1}{2} + m), \qquad \qquad \omega_{BTZ} = - k - i r_+ (2n + \tfrac{3}{2} + m) .
\end{equation}

For the Dirac field propagating in UAO black hole its QNF (\ref{e: QNF Dirac 1 AdS(2)}) and (\ref{e: QNF Dirac 2 AdS(2)}) coincide with the limit $k \to 0$ of the QNF (\ref{e: QNF BTZ Dirac}) for the same field moving in static three-dimensional BTZ black hole, which is different from the behavior of the Klein-Gordon field that we find in the previous Subsection. These facts illustrate that even in two dimensions the Klein-Gordon and Dirac fields react in a different way to the gravitational fields, as explained in Refs.\ \cite{Becar:2007hu}--\cite{LopezOrtega:2011sc}.

In a similar way to the QNF of the massive Klein-Gordon field, for the QNF (\ref{e: QNF Dirac 1 AdS(2)}) and (\ref{e: QNF Dirac 2 AdS(2)}) of the Dirac field, we obtain that $\im\, \omega < 0$ since $ m + \tfrac{1}{2} + n > 0$ and $m + \tfrac{3}{2} + n > 0$. Therefore in UAO black hole the Dirac field has stable QNM.

For the UAO black hole its QNF (\ref{e: QNF AdS(2) KG}), (\ref{e: QNF Dirac 1 AdS(2)}), and (\ref{e: QNF Dirac 2 AdS(2)}) scale linearly with the horizon radius $r_+$. This behavior is similar to that already found in static BTZ black hole \cite{Birmingham:2001pj,Cardoso:2001hn,Birmingham:2001hc} and large Schwarzschild anti-de Sitter black holes \cite{Horowitz:1999jd}--\cite{Cardoso:2003cj}. The Hawking temperature of the UAO black hole is \cite{Fabbri Navarro book}
\begin{equation} \label{e: Hawking temperature UAO}
 T_H = \frac{\hbar r_+}{2 \pi},
\end{equation}
where $\hbar$ is the reduced Planck constant. In units where $\hbar = 1$ we obtain that the QNF (\ref{e: QNF AdS(2) KG}), (\ref{e: QNF Dirac 1 AdS(2)}), and (\ref{e: QNF Dirac 2 AdS(2)}) are proportional to the Hawking temperature of the UAO black hole.

In Ref.\ \cite{LopezOrtega:2011sc} it is shown that in asymptotically flat two-dimensional black holes the solutions of the massless Dirac equation do not satisfy the boundary conditions of the QNM. Hence the massless Dirac field does not have well defined QNF in these black holes. In a similar way for asymptotically anti-de Sitter two-dimensional black holes we can show that the massless Dirac equation simplifies to the pair of differential equations for the radial functions $R_1$ and $R_2$ (see Eqs.\ (23) of \cite{LopezOrtega:2011sc})
\begin{align} \label{e: radial Dirac}
\frac{\dd R_{2}}{\dd r_*} + i\omega R_{2} =0, \qquad \qquad
\frac{\dd R_{1}}{\dd r_*} -i\omega R_{1} =0 .
\end{align}
We note that the solutions of these equations cannot satisfy the boundary conditions of the QNM for asymptotically anti-de Sitter two-dimensional black holes. Hence the massless Dirac field does not have well defined QNF in these black holes.

Furthermore for the UAO black hole we can show that in the massless limit the solutions of the radial equations (\ref{e: radial Dirac 1 2}) involve Jacobi polynomials \cite{Szego-book}. Analyzing these solutions we conclude that they cannot satisfy the boundary conditions of the QNM. Therefore we assert that the massless Dirac field does not have well defined QNF in this black hole.

\subsection{Time Times Temperature bound}
\label{s: TTT bound}

Studying how a perturbed thermodynamic system returns to an equilibrium state and considering thermodynamics and information theory, in Ref.\ \cite{Hod:2006jw} Hod proposes that the relaxation time $\tau$ of a perturbed thermodynamic system satisfies
\begin{equation} \label{e: TTT bound}
 \tau \geq \tau_{min} = \frac{\hbar}{\pi T},
\end{equation}
with $T$ denoting the temperature of the system. Thus a perturbed thermodynamic system has at least one relaxation mode that satisfies the inequality (\ref{e: TTT bound}). Hod calls to the bound (\ref{e: TTT bound}) as Time Times Temperature bound or briefly TTT bound \cite{Hod:2006jw}. See also Refs.\ \cite{Ropotenko:2007jm,Pesci:2008zv} for related work.

Furthermore Hod proposes that strong self-gravity objects, as the black holes, are the appropriate systems to test the TTT bound (\ref{e: TTT bound}). For these gravitational objects it is possible to show that the lower bound (\ref{e: TTT bound}) transforms into an upper bound on the absolute value of the imaginary part for the fundamental QNF (in this Subsection this quantity is denoted by $\omega_F$). Thus Hod finds that for a black hole the bound (\ref{e: TTT bound}) can be written as \cite{Hod:2006jw}
\begin{equation} \label{e: Hod bound}
 \omega_F \leq \frac{\pi T_H}{\hbar} ,
\end{equation}
where $T_H$ is Hawking's temperature. Defining the quantity $\mathbb{H} = \hbar \omega_F / (\pi T_H),$ we obtain that the bound (\ref{e: Hod bound}) becomes $\mathbb{H} \leq 1$.

The upper bound (\ref{e: Hod bound}) is valid for the fundamental QNF of several black holes as Schwarzschild, small Schwarzschild anti-de Sitter, Schwarzschild de Sitter, (extreme) Kerr, and Kerr-Newman \cite{Hod:2006jw,Hod:2008zz,Hod:2008se}.  Other gravitational systems saturate the upper bound (\ref{e: Hod bound}), for example, the Nariai spacetime \cite{LopezOrtega:2010dv} and it is expected that in the extremal limit the Schwarzschild de Sitter and the Kerr-Newman black holes saturate this bound \cite{Hod:2008se,Hod:2007tb}. Nevertheless in Ref.\ \cite{LopezOrtega:2010dv} it is shown that for the $D$-dimensional massless topological black hole, the three-dimensional BTZ black hole, and the $D$-dimensional de Sitter spacetime their fundamental QNF do not satisfy the upper bound (\ref{e: Hod bound}) (see also Ref.\ \cite{LopezOrtega:2010uu}).

We believe that for the UAO black hole it is convenient to use the QNF that we have calculated previously to test the upper bound (\ref{e: Hod bound}). For the massive Klein-Gordon field from the QNF (\ref{e: QNF AdS(2) KG}) we find that the frequency of the least damped QNM is equal to
\begin{equation} \label{e: fundamental QNM KG}
 \omega = - i r_+(\tfrac{1}{2} + \tfrac{1}{2}\sqrt{1+4m^2}) ,
\end{equation}
and therefore we obtain that for the Klein-Gordon field the quantity $\mathbb{H}$ is equal to $\mathbb{H}_{KG} = 1 + \sqrt{1+4m^2} > 1.$ Therefore in UAO black hole the fundamental QNF of the massive Klein-Gordon field does not satisfy the upper bound (\ref{e: Hod bound}).

From the QNF (\ref{e: QNF Dirac 1 AdS(2)}) and (\ref{e: QNF Dirac 2 AdS(2)}) for the Dirac field, we obtain that its least damped QNM has a frequency given by
\begin{equation}
 \omega = - ir_+ (m + \tfrac{1}{2}),
\end{equation}
and for this QNF of the Dirac field we find that the quantity $\mathbb{H}$ is equal to $\mathbb{H}_{D} = 1 + 2 m . $ In the previous Subsection we show that the QNF of the Dirac field are well defined for $m \neq 0$, and hence we obtain $\mathbb{H}_{D} > 1$ for the fermion field. Therefore in UAO black hole the fundamental QNF of the Dirac field does not fulfill the upper bound (\ref{e: Hod bound}). Thus from the expressions for $\mathbb{H}_{KG}$ and $\mathbb{H}_{D}$, we find that the bound (\ref{e: Hod bound}) is not valid in two-dimensional UAO black hole.

\section{Quasinormal frequencies of the CAO black hole}
\label{s: Charged AO black hole}

A generalization of the two-dimensional UAO black hole is the CAO black hole whose line element is \cite{Achucarro:1993fd}
\begin{equation} \label{e: metric charged AO black hole}
 \dd s^2 = \left( -M + \frac{r^2}{l^2} + \frac{J^2}{4 r^2} \right) \dd t^2 - \left( -M + \frac{r^2}{l^2} + \frac{J^2}{4 r^2} \right)^{-1} \dd r^2,
\end{equation}
where $M$ and $l$ are defined as for UAO black hole. In contrast to the UAO spacetime (\ref{e: metric AdS(2) black hole}) the CAO black hole has an event horizon and an inner horizon. In a similar way to the UAO black hole, the CAO black hole can be obtained from the dimensional reduction of the three-dimensional rotating BTZ black hole \cite{Banados:1992wn,Banados:1992gq,Carlip:1995qv} for which the quantity $J$ is its angular momentum and for the CAO black hole  $J$ is interpreted as a charge. It is straightforward to show that the CAO black hole is a solution to the equations of motion derived from the action \cite{Achucarro:1993fd}
\begin{equation}
 S = \int \dd^2 \sqrt{|g|} \,\, \phi \left( \mathcal{R} + \frac{2}{l^2} - \frac{J}{2 \phi^4} \right).
\end{equation}

We may expect that the QNF of the Klein-Gordon field moving in CAO black hole can be calculated with the method used to find the QNF of the same field propagating in spinning BTZ black hole \cite{Birmingham:2001pj,Birmingham:2001hc}. Nevertheless we find an obstacle. To understand it we compare the radial differential equations of the Klein-Gordon field in rotating BTZ and CAO black holes.

In Ref.\ \cite{Birmingham:2001hc} Birmingham shows that in spinning BTZ black hole \cite{Banados:1992wn,Banados:1992gq,Carlip:1995qv}
\begin{equation} \label{e: metric BTZ black hole}
 \dd s^2 = \left( -M + \frac{r^2}{l^2} + \frac{J^2}{4 r^2} \right) \dd t^2 - \left( -M + \frac{r^2}{l^2} + \frac{J^2}{4 r^2} \right)^{-1} \dd r^2 - r^2 \left( \dd \phi - \frac{J}{2r^2} \dd t \right)^2,
\end{equation}
the Klein-Gordon equation (\ref{e: Klein Gordon}) simplifies to a radial differential equation when we propose the separable solution
\begin{equation} \label{e: ansatz BTZ rotating}
 \Phi = R_{BTZ}(r) \textrm{e}^{- i \omega t} \textrm{e}^{i k \phi},
\end{equation}
where $k$ is the azimuthal number (in Ref.\ \cite{Birmingham:2001hc} it is denoted by $m$).

Denoting by $r_+$ and $r_-$ the radii of the event horizon and the inner horizon of the BTZ black hole (or of the CAO black hole) we find that these are equal to
\begin{equation} \label{e: roots horizon BTZ}
 r_\pm^2 = \frac{l^2}{2} \left( M \pm \sqrt{M^2 - \frac{J^2}{l^2}} \right),
\end{equation}
and using the variable\footnote{Taking the limit $r_- \to 0$ of the variable $z$ in the expression (\ref{e: z variable}) we get the variable $z$ of the formula (\ref{e: alpha beta Klein-Gordon}). }
\begin{equation} \label{e: z variable}
 z = \frac{r^2-r_+^2}{r^2-r_-^2} ,
\end{equation}
Birmingham finds that in spinning BTZ black hole the radial differential equation for the Klein-Gordon field is equal to (see Eqs.\ (8) and (9) in Ref.\ \cite{Birmingham:2001hc})
\begin{equation} \label{e: radial BTZ Klein-Gordon z}
 z (1-z) \frac{\dd^2 R_{BTZ}}{\dd z^2} + (1-z) \frac{\dd R_{BTZ}}{\dd z} + \left( \frac{A}{z} + B + \frac{C}{1-z} \right) R_{BTZ} = 0 ,
\end{equation}
where
\begin{align} \label{e: A B C BTZ Klein-Gordon}
A = \frac{l^4 \left( \omega r_+ -\frac{k r_-}{l} \right)^2}{4(r_+^2 -r_-^2)^2} , %\nonumber \\
\quad B = -\frac{l^4 \left( \omega r_- -\frac{k r_+}{l} \right)^2}{4(r_+^2 -r_-^2)^2} , %\\
\quad C = - \frac{m^2 l^2}{4} .%\nonumber
\end{align}
In a similar way to the variable $z$ of the formula (\ref{e: alpha beta Klein-Gordon}), we find that the variable of the expression (\ref{e: z variable}) behaves in the form, $z=0$ as $r=r_+$ and $z \to 1$ as $r \to \infty$.

In Ref.\ \cite{Birmingham:2001hc} it is shown that the solutions of Eq.\ (\ref{e: radial BTZ Klein-Gordon z}) involve hypergeometric functions and hence it is possible an exact calculation of the QNF for the massive Klein-Gordon field propagating in spinning BTZ black hole (see also \cite{Birmingham:2001pj}).

For the CAO black hole (\ref{e: metric charged AO black hole}), if we propose a separable solution as that of the formula (\ref{e: separation KG}), then we obtain that the Klein-Gordon equation (\ref{e: Klein Gordon}) simplifies to the radial differential equation
\begin{equation}
 \frac{\dd }{\dd r}\left( \left( -M + \frac{r^2}{l^2} + \frac{J^2}{4 r^2} \right) \frac{\dd R}{\dd r}  \right) + \left( -M + \frac{r^2}{l^2} + \frac{J^2}{4 r^2} \right)^{-1} \omega^2 R - m^2 R = 0 ,
\end{equation}
and using the variable $z$ of the formula (\ref{e: z variable}) we obtain that this equation transforms into
\begin{equation} \label{e: radial AO Klein-Gordon z}
 z (1-z) \frac{\dd^2 R}{\dd z^2} + \left(1-z -  \frac{r_+^2-r_-^2}{2} \frac{z}{r_+^2 - z r_-^2}\right) \frac{\dd R}{\dd z} + \left( \frac{\alpha}{z} + \gamma  + \frac{ \beta }{1-z} \right) R = 0 ,
\end{equation}
where\footnote{We point out that the parameter $ \alpha$  of the CAO black hole is different from that of the formula (\ref{e: alpha beta Klein-Gordon}) for the UAO black hole. The parameter $\beta$ is equal for both black holes. }
\begin{align}
 \alpha = \frac{l^4 \omega^2r_+^2}{4 (r_+^2-r_-^2)^2}  = \lim_{k \to 0} A, %\nonumber \\
 \quad \gamma = - \frac{l^4 \omega^2r_-^2}{4 (r_+^2-r_-^2)^2}  =  \lim_{k \to 0} B , %\\
\quad \beta  = - \frac{m^2 l^2}{4} = C.  %\nonumber
\end{align}
Notice that the parameters $\alpha$ and $\gamma$ coincide with the limit $k \to 0$ of the quantities $A$ and $B$ of the formula (\ref{e: A B C BTZ Klein-Gordon}) for the spinning BTZ black hole and it is valid that $\beta = C$. For $r_-=0$, Eq.\ (\ref{e: radial AO Klein-Gordon z}) reduces to the radial equation (\ref{e: radial Klein Gordon}) of the UAO black hole.

The relevant fact to our work is that the radial equation (\ref{e: radial AO Klein-Gordon z}) of the CAO black hole has an additional singular point at $z=r_+^2 /r_-^2$ compared with the radial equation (\ref{e: radial BTZ Klein-Gordon z}) of the spinning BTZ black hole. It is due to the last term in the factor that multiplies to the first derivative of the radial function $R$. As a consequence, to solve Eq.\ (\ref{e: radial AO Klein-Gordon z}) the method of Ref.\ \cite{Birmingham:2001hc} does not work and we have not be able to obtain exact solutions in terms of hypergeometric functions. Surprisingly, the radial differential equation for the two-dimensional CAO black hole is more complicated than the radial equation of the three-dimensional rotating BTZ black hole. We believe that this result is due (at least in part) to the fact that the spinning BTZ black hole is a constant curvature spacetime \cite{Banados:1992wn}--\cite{Carlip:1995qv}, whereas the scalar curvature of the CAO black hole is equal to \cite{Achucarro:1993fd}
\begin{equation} \label{e: scalar curvature AO}
 \mathcal{R} = - \frac{2}{l^2} - \frac{3 J}{2 r^4} ,
\end{equation}
and it is not constant for $J \neq 0$.

As a consequence for the CAO black hole it is not easy to extend the results for the QNF of the massive Klein-Gordon field to the field non-minimally coupled to the scalar curvature, since it is not constant (see the formula (\ref{e: scalar curvature AO})). In contrast, in the three-dimensional BTZ black hole it is straightforward to extend the results on the QNF of the minimally coupled massive Klein-Gordon field to the field non-minimally coupled to the scalar curvature, because the spacetime is of constant curvature, and in its equation of motion the term proportional to scalar curvature only produces a shift in the mass term. Thus we believe that in CAO black hole this problem is more complicated than in spinning BTZ black hole.

For these reasons, in what follows we calculate numerically the QNF of the minimally coupled massive Klein-Gordon field, since we expect that the massless Klein-Gordon field does not have well defined QNF (see Subsection \ref{s: QNF Klein-Gordon} and Ref.\ \cite{LopezOrtega:2011sc}).

\subsection{Numerical method}

To calculate numerically the QNF of the massive Klein-Gordon field in CAO black hole we exploit the method proposed by Horowitz and Hubeny \cite{Horowitz:1999jd} that allows us to compute numerically the QNF of asymptotically anti-de Sitter black holes (see also \cite{Berti:2009kk,Konoplya:2011qq,Cardoso:2001bb,Cardoso:2003cj,Konoplya:2002zu,Wang:2000gsa,Berti:2003ud,Cardoso:2001hn}). In the following we expound briefly the method and point out some  relevant facts for computing the QNF of the Klein-Gordon field moving in CAO black hole.

For a $D$-dimensional spherically symmetric spacetime of the form
\begin{equation}
 \dd s^2 = f \dd t^2 - \frac{\dd r^2}{f} - r^2 \dd \Omega^2_{D-2},
\end{equation}
where $f$ is a function of the coordinate $r$ and $\dd \Omega^2_{D-2}$ is the line element of a $(D-2)$-dimensional sphere, it is well known that the equations of motion for perturbation fields simplify to Schr\"odinger type equations \cite{Kokkotas:1999bd}--\cite{Konoplya:2011qq}
\begin{equation} \label{e: Schrodinger equation}
 \frac{\dd^2 U}{\dd r_*^2} +(\omega^2 - V)U = 0,
\end{equation}
where $V$ is an effective potential that depends on the type of perturbation and $U$ is related to the radial function.

Assuming a time dependence of the form $\exp{(-i \omega t)}$ (as previously) and imposing that the function $U$  near the horizon satisfies the purely ingoing boundary condition of the QNM, gives as a consequence the following form for this function
\begin{equation}
 U(r_*) = \exp{(-i \omega r_*)} U_1(r_*) .
\end{equation}
From Eq.\ (\ref{e: Schrodinger equation}) we find that the function $U_1$ satisfies \cite{Berti:2009kk,Horowitz:1999jd}
\begin{equation} \label{e: equation function u}
 f \frac{\dd^2 U_1}{\dd r^2} + \left( \frac{\dd f}{\dd r} - 2 i \omega \right)\frac{\dd U_1}{\dd r} - \frac{V}{f} U_1 = 0.
\end{equation}

For asymptotically anti-de Sitter spacetimes we impose the second boundary condition of the QNM as $r \to \infty$ (the field decays to zero as $r \to \infty$). To deal with it, Horowitz and Hubeny make the change of variable \cite{Horowitz:1999jd}
\begin{equation} \label{e: change of variable x r}
 x=\frac{1}{r},
\end{equation}
in such a way that they impose the second boundary condition at $x=0$.

Making this change of variable, we find that Eq.\ (\ref{e: equation function u}) transforms into
\begin{equation} \label{e: differential equation x}
 s \frac{\dd^2 U_1}{\dd x^2} + \frac{t}{x-x_+} \frac{\dd U_1}{\dd x} + \frac{u}{(x-x_+)^2} U_1 = 0,
\end{equation}
where $x_+=1/r_+$, and we determine the expressions for  $s$, $t$, $u$ from the effective potential $V$ and the function $f$. If the singular point at $x=x_+$ is regular, then we can expand the function $U_1$ in a power series of the form
\begin{equation} \label{e: power series u}
 U_1= (x-x_+)^\nu \sum_{k=0}^{\infty} a_k (\omega) (x-x_+)^k ,
\end{equation}
where the boundary conditions at the event horizon fix the exponent $\nu$. Moreover we expand the function $u$ in a power series as follows
\begin{equation} \label{e: u Taylor series}
 u =  \sum_{k=0}^{\infty} u_k (x-x_+)^ k,
\end{equation}
and similarly for the functions $s$ and $t$.

Substituting the power series of $U_1$, $s$, $t$, and $u$ in the differential equation (\ref{e: differential equation x}), we find the following recurrence relation for the coefficients $a_k(\omega)$ \cite{Horowitz:1999jd}
\begin{equation} \label{e: recurrence relation}
 a_k = -\frac{1}{k(k-1)s_0 + k t_0 } \sum_{n=0}^{k-1} a_n ( n(n-1) s_{k-n} + n t_{k-n} + u_{k-n}).
\end{equation}
The boundary condition at infinity ($x=0$) imposes that  $ U_1 =0$, hence we get that the QNF are the roots of \cite{Horowitz:1999jd}
\begin{equation} \label{e: equation for QNF}
  \sum_{k=0}^{\infty} a_k (\omega) (-x_+)^k=0 .
\end{equation}

Usually to solve Eq.\ (\ref{e: equation for QNF}) we take a partial sum of degree $N$ and find the roots of the resulting polynomial. To verify the obtained results for the QNF we calculate the roots of some higher order partial sums and if the roots converge, then the method is reliable. This method is widely used to determine the QNF of the asymptotically anti-de Sitter black holes, (see for example Refs.\ \cite{Horowitz:1999jd,Cardoso:2001bb,Cardoso:2003cj,Konoplya:2002zu,Wang:2000gsa,Berti:2003ud,Cardoso:2001hn} and the reviews \cite{Berti:2009kk,Konoplya:2011qq} for additional examples).

To use this method for computing the QNF of the Klein-Gordon field in CAO black hole we note the following points. For the two-dimensional metric of the form (\ref{e: metric general}) with metric functions $f$ and $g$, in Ref.\ \cite{LopezOrtega:2011sc} it is shown that the massive Klein-Gordon equation (\ref{e: Klein Gordon}) simplifies to a Schr\"odinger type equation (\ref{e: Schrodinger equation}) with an effective potential equal to
\begin{equation}
 V = m^2 f.
\end{equation}

Following a similar procedure to that explained in Ref.\ \cite{Horowitz:1999jd} we obtain that for the two-dimensional spacetimes with metric (\ref{e: metric general}) the differential equation analogous to (\ref{e: equation function u}) is
\begin{equation} \label{e: equation function u f g}
g \frac{\dd^2 U_1}{\dd r^2} + \left( \frac{f \frac{\dd g}{\dd r} + g \frac{\dd f}{\dd r} }{2 f} - 2 i \omega \sqrt{\frac{g}{f}} \right) \frac{\dd U_1}{\dd r} -m^2  U_1= 0 .
\end{equation}
If the functions $f$ and $g$ satisfy $f = g$, as for the CAO black hole, Eq.\ (\ref{e: equation function u f g}) simplifies to
\begin{equation} \label{e: equation function u two-dimensional}
 f \frac{\dd^2 U_1}{\dd r^2} + \left( \frac{\dd f}{\dd r} - 2 i \omega \right)\frac{\dd U_1}{\dd r} - m^2 U_1 = 0 .
\end{equation}

In what follows we choose units where $l=1$ and for the CAO black hole it is convenient to write the function $f$ as
\begin{equation}
 f = \frac{(r^2 -r_+^2)(r^2-r_-^2)}{r^2},
\end{equation}
where the horizon radii $r_+$ and $r_-$  are defined in the formula (\ref{e: roots horizon BTZ}). Using the variable $x$ we get that the previous function becomes
\begin{equation}
 f = \frac{(x^2 - x_+^2)(x^2-x_-^2)}{x^2 x_-^2 x_+^2} ,
\end{equation}
with $x_-=1/r_-$.

In this spacetime we obtain that the differential equation (\ref{e: equation function u two-dimensional}) transforms into an equation of the form (\ref{e: differential equation x}) with the functions $s$, $t$, and $u$ equal to
\begin{align}
 s &= \frac{x^2 (x + x_+)(x^2 - x_-^2) }{x_+^2 x_-^2}, \quad \qquad u = -m^2 (x-x_+), \nonumber \\
t &= \frac{2 x (x^2 - x_+^2)(x^2 - x_-^2)}{x_+^2 x_-^2} + \frac{2 x^5}{x_+^2 x_-^2} + 2 i \omega x^2 -2 x .
\end{align}
We note that the resulting differential equation has regular singular points at $x=0,\,\, x_-$, $ x_+$. To work with a non-extremal CAO black hole we must impose the condition
\begin{equation} \label{e: inequality first}
r_+ > r_- \quad \iff \quad  x_- > x_+ .
\end{equation}

We must consider that the power series (\ref{e: power series u}) has a radius of convergence as large as the distance to the nearest singular point of the differential equation. In contrast to the Schwarzschild anti-de Sitter spacetime, the CAO black hole has one inner horizon radius and therefore we must be careful about the singular point located at this horizon. In order that the convergence radius of the power series (\ref{e: power series u}) reaches the point $x=0$ we must impose that the distance between the singular points $x_+$ and $x_-$ is larger than the distance between the singular points $x_+$ and $0$, that is,
\begin{equation} \label{e: inequality second}
 x_- - x_+ > x_+ - 0 \,\, \iff \,\,  x_-> 2 x_+ \,\, \iff \,\, r_+ > 2 r_-.
\end{equation}

%%%%%%%%%%%%%%%%%%%%%%%%%%%%%%%%%%%%%%%%%%%%%%%%%%%%%%%%%%%%%%%%%%%%%%%%%%%%%%%%%%%%%%%%%%%%%%%%%%%

\begin{figure}[htbp]
\begin{center}
\includegraphics[clip,scale=.67]{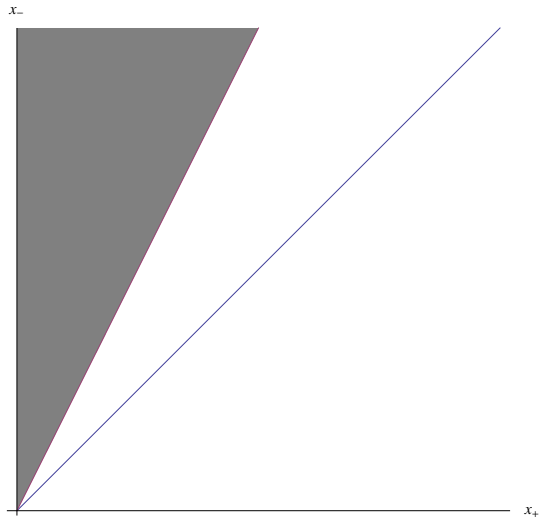}
\caption{The (non)-extreme CAO black holes are located (above) on the line $x_-=x_+$. The CAO black holes whose parameters fall in the shaded region satisfy the inequality (\ref{e: inequality second}) and therefore the Horowitz-Hubeny method works for the black holes of this region.}
\label{fig: parameter region}
\end{center}
\end{figure}

If we satisfy the inequality (\ref{e: inequality second}), then we automatically fulfill the inequality (\ref{e: inequality first}). Therefore  in CAO black hole, to determine the QNF of the Klein-Gordon field with the Horowitz-Hubeny method we must restrict its parameters in such a way that we satisfy the inequality (\ref{e: inequality second}). Using the event horizon radius $r_+$ and the charge $J$ we write the inequality (\ref{e: inequality second}) in the form
\begin{equation} \label{e: inequality equivalent}
 r_+^2 > J .
\end{equation}

If we do not fulfill the inequality (\ref{e: inequality second}) (or (\ref{e: inequality equivalent})) we cannot ensure that the convergence radius of the power series (\ref{e: power series u}) includes to the point $x=0$. Thus we must restrict  our search of QNF to the CAO black holes whose parameters fall in the shaded region of Fig. \ref{fig: parameter region}. Something similar happens in four-dimensional Reissner-Nordstr\"om anti-de Sitter black hole, for which it is known that the Horowitz-Hubeny method does not work in the extremal limit \cite{Wang:2000gsa,Berti:2003ud}.

In the previous section we show that the QNF of the asymptotically anti-de Sitter two-dimensional black holes are not defined in the massless limit. Doubtless it is convenient to determine what happens with our numerical method in this limit.

We note that in the massless limit our numerical method does not work because the power series $\sum_{k=0}^\infty a_k (\omega) (x-x_+)^k$ of the formula (\ref{e: power series u}) goes to a constant. To see this fact we point out that for the function $u$ of the formula (\ref{e: u Taylor series}) its Taylor expansion has the property that only the coefficient $u_1$ is different from zero. Thus from the recurrence relation (\ref{e: recurrence relation}) we get that the first two coefficients $a_k$ are
\begin{align}
 a_1 & = -\frac{m^2}{-2 x_+ + 2x_+^3/x_-^2 + 2 i x_+^2 \omega} ,\nonumber \\
a_2 & = \frac{m^2 (-16 + 32x_+^2/x_-^2 + 4 i \omega x_+ - m^2)}{(-4x_+ + 4x_+^3/x_-^2 + 2i \omega x_+^2)(-2x_+ + 2x_+^3 /x_-^2 + 2 i \omega x_+^2 )} .
\end{align}
From these expressions we note that both coefficients are proportional to $m^2$ and therefore in the massless limit we find $a_1=a_2=0$.

For the coefficient $a_3$ we get
\begin{equation}
 a_3 = - \frac{1}{6 s_0 + 3 t_0} [a_1 t_2 + a_2 (2 s_1 + 2 t_1 + u_1)],
\end{equation}
and we notice that there is no term proportional to $a_0$ because $u_k = 0$ for $k \geq 2$.  Since $a_1$ and $a_2$ are proportional to $m^2$, we find that $a_3$ is also proportional to $m^2$. Hence in the massless limit we obtain $a_3 = 0$.

Using these results we can assume that $a_1, a_2, \ldots , a_p$ are proportional to $m^2$ and determine the behavior of $a_{p+1}$ in the massless limit. From the recurrence relation (\ref{e: recurrence relation}) we find that $a_{p+1}$ is equal to
\begin{equation}
 a_{p+1} = - \frac{a_1 t_p + a_2 (2 s_{p-1} + 2 t_{p-1} + u_{p-1}) + \ldots + a_p ((p-1)p s_1 + p t_1 + u_1)}{p(p+1) s_0 + (p+1) t_0 }.
\end{equation}
From this expression, since $a_{p+1}$ is proportional to $a_1, a_2, \ldots , a_p$, we find that the coefficient $a_{p+1}$ is proportional to $m^2$ and hence it goes to zero in the massless limit.

Thus we obtain that in the massless limit the power series of the formula  (\ref{e: power series u}) goes to a constant and therefore our numerical method does not work. It is consistent with the expected absence of QNF in the massless limit.

\subsection{Numerical results}

To find the QNF of the massive Klein-Gordon field moving in CAO black hole we use the method by Horowitz and Hubeny. For the Klein-Gordon field propagating in CAO black hole in our computations we always obtain that its QNF are purely imaginary, thus we expect that the QNM of the CAO black hole are purely damped, as the QNM of the UAO black hole previously calculated in Section \ref{s: AdS(2) black hole}. Moreover in our numerical computations for the QNF of the Klein-Gordon field  we find that their imaginary parts satisfy $\im \,\omega < 0$, hence in CAO black hole the QNM of the Klein-Gordon field are stable.

%%%%%%%%%%%%%%%%%%%%%%%%%%%%%%%%%%%%%%%%%%%%%%%%%%%%%%%%%%%%%%%%%%%%%%%%%%%%%%%%%%%%%%%%%%%%%%%%%%

\begin{figure}[htbp]
\begin{center}
\includegraphics[clip]{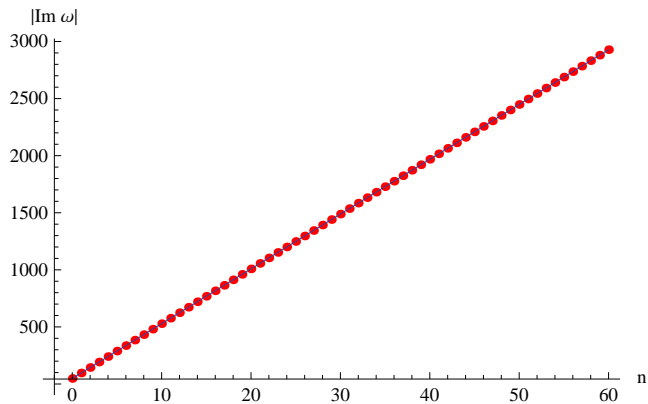}
\caption{For the CAO black hole with horizon radii $r_+=50$, $r_-=10$, and the Klein-Gordon field of mass $m=1/10$, we plot the absolute values of the imaginary parts for their QNF as a function of the mode number $n$. In this Figure we draw the QNF with mode numbers $n=0,1,\dots,60$. }
\label{fig: sixty modes}
\end{center}
\end{figure}

First we analyze the change of the imaginary parts of the QNF with the mode number. In Fig.\ \ref{fig: sixty modes} we plot the absolute values of the imaginary parts for sixty one QNF of the Klein-Gordon field with mass $m=1/10$ propagating in a CAO black hole with horizon radii $r_+ = 50$, $r_-=10$. We see in Fig.\ \ref{fig: sixty modes} that there is a linear relation between $|\im \, \omega |$ and the mode number $n$. For the points of Fig.\ \ref{fig: sixty modes} the linear fit is
\begin{equation}
 |\im \, \omega | = 48.47 + 48.00 \, n .
\end{equation}
This behavior is similar to that for the QNF of the Klein-Gordon, electromagnetic, and gravitational perturbations propagating in large Schwarzschild anti-de Sitter spacetime \cite{Horowitz:1999jd,Cardoso:2003cj}. For CAO black holes with  other values of $r_+$ and $r_-$ for which we compute their QNF we also find similar linear relations between $|\im \, \omega |$ and $n$.

Motivated by some conjectures \cite{Hod:1998vk,Maggiore:2007nq}, recently there is interest in the computation of the asymptotic QNF. For the CAO black hole we calculate the QNF for $n$ sufficiently large and from our numerical results we note that there is a mode number $n_1$ depending on $r_+$ and $r_-$, such that for $n > n_1$ the asymptotic QNF are equally spaced (see some representative asymptotic QNF in Table \ref{t: AQNF CAO}). Our numerical results display that this behavior does not change as the mode number $n$ increases even more. (In some cases we have calculated QNF with mode numbers as large as $n=250$.)

%%%%%%%%%%%%%%%%%%%%%%%%%%%%%%%%%%%%%%%%%%%%%%%%%%%%%%%%%%%%%%%

\begin{table}[ht]
\caption{Asymptotic QNF of the CAO black hole for different values of the horizon radii $r_+$ and $r_-$.}
\label{t: AQNF CAO}
\begin{center}
\begin{tabular}{|c|c|c|c|}
\hline
\,\,$n$\,\, & \,\,$\omega$\,\, & \,\,$n$\,\, & \,\,$\omega$ \,\,\\ \hline
\multicolumn{4}{|c|}{$r_+ = 50$,  $r_- = 10$ \vline \,\,\,\,  $r_+ = 50$, $r_- = 20$} \\ \hline
91 & -4416.00 $i$ & 122 & -5166.00 $i$ \\ \hline
92 & -4464.00 $i$ & 123 & -5208.00 $i$\\ \hline
93 & -4512.00 $i$ & 124 & - 5250.00 $i$ \\ \hline
94 & -4560.00 $i$ & 125 & -5292.00 $i$ \\ \hline
\multicolumn{4}{|c|}{$r_+ = 100$, $r_- = 10$  \vline \,\,\,\,\,\, $r_+ = 25$, $r_- = 10$} \\ \hline
109 & -10890.00 $i$ & 132  & -2793.00 $i$  \\ \hline
110 & -10989.00 $i$ & 133  & -2814.00 $i$ \\ \hline
111 & -11088.00 $i$ & 134  & -2835.00 $i$ \\ \hline
112 & -11187.00 $i$ & 135  & -2856.00 $i$ \\ \hline
\end{tabular}
\end{center}
\end{table}

%%%%%%%%%%%%%%%%%%%%%%%%%%%%%%%%%%%%%%%%%%%%%%%%%%%%%%%%%%%%%%%

From the numerical values for the asymptotic QNF showed in Table \ref{t: AQNF CAO} we find that the spacing of the asymptotic QNF is well fitted by the formula
\begin{equation}
 \Delta \omega = \omega_{n+1} - \omega_n = -  \frac{r_+^2 - r_-^2}{r_+} i = - \kappa i ,
\end{equation}
where $\kappa = (r_+^2 - r_-^2)/r_+$ is the surface gravity. Notice that for the three-dimensional spinning BTZ black hole the spacing of the asymptotic QNF  is determined by\footnote{In units where $l=1$, because in our numerical computations we choose this value.} (from the formula (18) in \cite{Birmingham:2001hc})
\begin{equation}
 \Delta \omega_L = - 2(r_+ - r_-) i, \qquad \Delta \omega_R = - 2(r_+ + r_-) i .
\end{equation}
Hence for the CAO and BTZ black holes their asymptotic QNF behave in a different way.

%%%%%%%%%%%%%%%%%%%%%%%%%%%%%%%%%%%%%%%%%%%%%%%%%%%%%%%%%%%%%%%%%%%%%%%%%%%%%%%%%%%%%%%%%%%%%%%%%%

\begin{figure}[htbp]
\begin{center}
\includegraphics[clip]{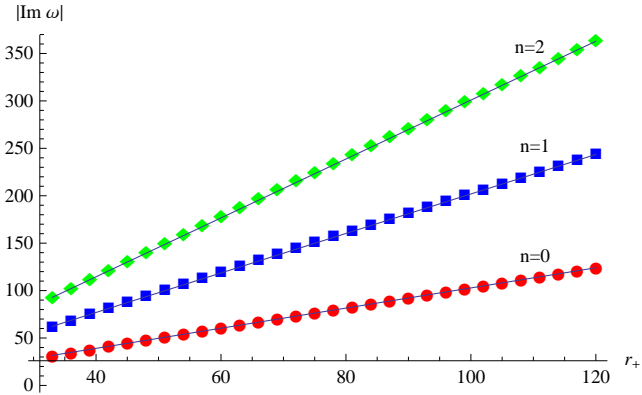}
\caption{For a Klein-Gordon field of mass $m=1/5$ moving in CAO black holes with constant inner horizons equal to $r_-=10$, we plot the absolute values of the imaginary parts for the QNF as a function of the horizon radius $r_+$. We take increments in the horizon radius $\Delta r_+= 3$. In the Figure we draw the QNF with mode number $n=0$ (circles), $n=1$, (squares), $n=2$ (diamonds).   }
\label{fig: variation r plus}
\end{center}
\end{figure}

For the first three QNF of the Klein-Gordon field with mass equal to $m=1/5$, moving in CAO black holes with inner radius $r_-=10$, in Fig.\ \ref{fig: variation r plus} we draw the absolute values of their imaginary parts for increasing values of the horizon radius $r_+$ and we point out that in this Figure the horizon radius $r_+$  varies in steps of $\Delta r_+ =3$. We observe in Fig.\ \ref{fig: variation r plus} that for fixed inner radius $r_-$ the relation between $|\im \, \omega |$ and $r_+$ is linear. For the three modes plotted in Fig.\ \ref{fig: variation r plus} the linear fits of the data are
\begin{eqnarray}\label{e: fits  r plus change}
 n= 0 : & |\im \,\,\, \omega | = -3.41 + 1.06\, r_+ , \nonumber \\
n= 1 : & |\im \,\,\, \omega | = -6.37 + 2.08 \, r_+ , \\
n= 2 : & |\im \,\,\, \omega | = -9.68 + 3.10 \, r_+ . \nonumber
\end{eqnarray}

In the linear fits (\ref{e: fits  r plus change}) the horizon radius fulfills $r_+ > 20$ to satisfy the inequality (\ref{e: inequality second}), thus for $n=0,1,2$ we see that the linear fits (\ref{e: fits  r plus change}) produce that $|\im \,\,\, \omega | > 0$. Notice that the slope of the linear relations (\ref{e: fits  r plus change}) depends on the mode number. This linear dependence recall us the formula (\ref{e: QNF AdS(2) KG}) for the QNF of the UAO black hole (and the formula (\ref{e: QNF BTZ Klein Gordon}) for the QNF of the static BTZ black hole) where we see a linear relation between the imaginary parts of the QNF and the horizon radius. For the CAO black holes with inner radius $r_- = 3$, 7, and 13 we also find linear relations similar to those of the formulas (\ref{e: fits  r plus change}) when the horizon radius $r_+$ increases.

\begin{figure}[htbp]
\begin{center}
\includegraphics[clip]{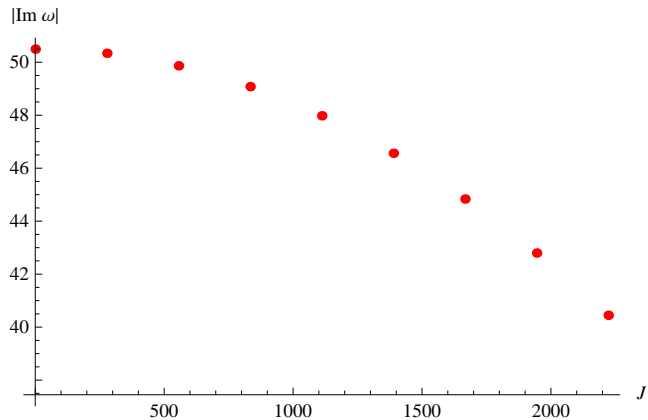}
\caption{For a Klein-Gordon field of mass $m=1/10$ and CAO black holes with horizon radius $r_+ = 50$ we plot the absolute values of the imaginary parts for the fundamental QNF as a function of the charge $J$. We draw the QNF for values of $J$ that satisfy the inequality (\ref{e: inequality equivalent}).}
\label{fig: fundamental mode}
\end{center}
\end{figure}

\begin{figure}[htbp]
\begin{center}
\includegraphics[clip]{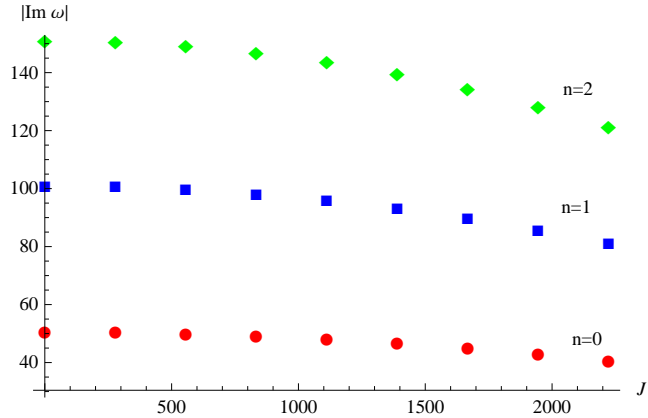}
\caption{For a Klein-Gordon field of mass $m=1/10$ and CAO black holes with $r_+ = 50$  we plot the absolute values of the imaginary parts for the first three QNF as a function of the charge $J$. Here we use circles to represent the $n=0$ mode, squares for the $n=1$ mode, and diamonds for the $n=2$ mode. We plot the QNF of black holes whose charge $J$ satisfies the inequality (\ref{e: inequality equivalent}).}
\label{fig: first three}
\end{center}
\end{figure}

For the Klein-Gordon field of mass equal to $m=1/10$, in Fig.\ \ref{fig: fundamental mode} we plot the absolute values of the imaginary parts for the fundamental modes of the CAO black holes with horizon radius $r_+ = 50$ and several values of the charge $J$ compatible with the inequality (\ref{e: inequality equivalent}). Furthermore for the same values of $r_+$ and $m$ in Fig.\ \ref{fig: first three} we draw the absolute values for the imaginary parts of the first three QNF as a function of the charge $J$. Notice that  in Figs.\ \ref{fig: fundamental mode} and \ref{fig: first three} the points on the ordinate axis are the absolute values of the imaginary parts for the QNF of the UAO black hole with horizon radius $r_+ = 50$.

From Figs.\ \ref{fig: fundamental mode} and \ref{fig: first three} we notice that for fixed $r_+$ and $m$ the absolute values of the imaginary parts of the QNF decrease as the charge $J$ increases, hence the decay times increase as the charge $J$ increases.  Thus we get that for the same values of the quantities $r_+$, $m$, and $n$ the QNM of the UAO black hole are more damped than those of the CAO black hole. This behavior is different from that for the four-dimensional Reissner-Nordstr\"om anti-de Sitter black hole. For this black hole, in the region of parameters explored in Refs.\ \cite{Wang:2000gsa,Berti:2003ud}, it is found that if the electric charge increases, then the absolute values of the imaginary parts for the QNF increase. In Fig.\ \ref{fig: first three} we also observe that for the same variation in the charge $J$, the change of $ | \im\, \omega | $ is greater when the mode number $n$ increases.

For the same value of the mass $m$ we also compute the QNF of the Klein-Gordon field moving in CAO black holes with $r_+=100$, 25, 15. For the absolute values of their imaginary parts we find similar behaviors to those of Figs.\ \ref{fig: fundamental mode} and \ref{fig: first three} for $r_+=50$, but if the horizon radius decreases, then the convergence of the numerical method is slower.

%%%%%%%%%%%%%%%%%%%%%%%%%%%%%%%%%%%%%%%%%%%%%%%%%%%%%%%%%%%%%%%%%%%%%%%%%%%%%%%%%%%%%%%%%%%%%%%%%%

\begin{figure}[htbp]
\begin{center}
\includegraphics[clip]{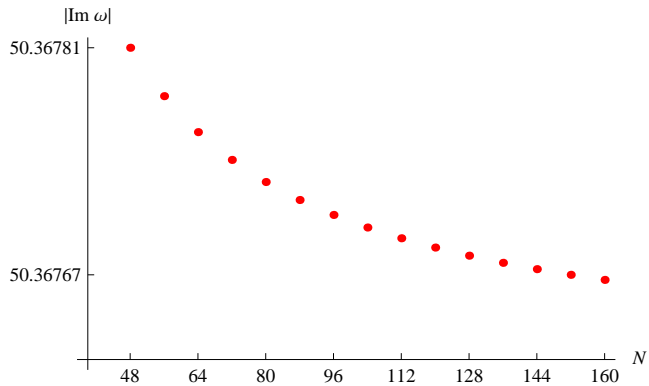}
\caption{Convergence curve up to polynomials of degree $N=160$ for the fundamental mode $n=0$ of the Klein-Gordon field with mass equal to $m=1/10$ in a CAO black hole with horizon radii $r_+=50$ and $r_-=5/2$. }
\label{fig: convergence curve}
\end{center}
\end{figure}

To obtain the QNF plotted in Figs.\ \ref{fig: variation r plus}, \ref{fig: fundamental mode}, and \ref{fig: first three}, in the Horowitz-Hubeny method we use polynomials of degree $N=120$. We believe that polynomials of this degree are appropriate for the determination of the QNF, since the convergence curves show that for $N \geq 100$ the first roots of the polynomials change slowly as the degree of the polynomials increases. See Fig.\ \ref{fig: convergence curve} for the convergence curve of the fundamental mode for the Klein-Gordon field of mass $m=1/10$ propagating in a CAO black hole with $r_+=50$ and $r_-= 5/2$.  The convergence curve plotted in Fig.\ \ref{fig: convergence curve} has a behavior similar to that of Fig.\ 1 in \cite{Govindarajan:2000vq}, but notice that in the previous reference a different numerical method is used for the calculation of the QNF.

Furthermore we compute the convergence curves of CAO black holes with $r_+=50$ and other values of the inner horizon radius compatible with the inequality (\ref{e: inequality equivalent}). We also calculate the convergence curves of CAO black holes with horizon radii equal to $r_+=100$, 25, 15. For these black holes we obtain that their convergence curves behave as the curve of Fig.\ \ref{fig: convergence curve}.

%%%%%%%%%%%%%%%%%%%%%%%%%%%%%%%%%%%%%%%%%%%%%%%%%%%%%%%%%%%%%%%%%%%%%%%%%%%%%%%%%%%%%%%%%%%%%%%%%%

%%%%%%%%%%%%%%%%%%%%%%%%%%%%%%%%%%%%%%%%%%%%%%%%%%%%%%%%%%%%%%%%%%%%%%%%%%%%%%%%%%%%%%%%%%%%%%%%%%

\section{Summary}
\label{s: Summary}

In UAO black hole we calculate exactly the QNF of the massive Klein-Gordon and Dirac fields, furthermore we show that for the massless Klein-Gordon and Dirac fields  their QNF are not well defined in this background. For the massive Klein-Gordon and Dirac fields moving in UAO black hole we obtain that their QNF are purely imaginary and for these fields we also find that their QNM are stable. Using the computed values of the QNF we prove that the fundamental frequency of the UAO black hole does not satisfy the upper bound (\ref{e: Hod bound}) derived from the TTT bound (\ref{e: TTT bound}). Thus our results support that the upper bound (\ref{e: Hod bound}) is not universal, as is asserted in Ref.\ \cite{LopezOrtega:2010dv}, and hence for gravitational systems we must determine its applicability limits.

For the CAO black hole we have not been able to calculate exactly the QNF of the massive Klein-Gordon field, because its radial equation has one additional singular point compared with the radial equation of the rotating BTZ black hole; hence we exploit the Horowitz Hubeny numerical method \cite{Horowitz:1999jd} to compute them. As for the UAO black hole we obtain purely imaginary QNF and they produce stable QNM. For the Klein-Gordon field and  the same values of the mass $m$  and the horizon radius $r_+$, the QNF of the UAO black hole are more damped than the QNF of the CAO black hole. For fixed values of the field mass $m$ and the  horizon radius $r_+$ we also find that in CAO black hole the decay times increase as the charge $J$ increases, in contrast to the four-dimensional Reissner-Nordstr\"om anti-de Sitter black hole, for which the decay times decrease as the electric charge increase \cite{Wang:2000gsa,Berti:2003ud}. Moreover from our numerical results we guess a formula for the spacing of the asymptotic QNF for the CAO black hole. 

To define the QNM of asymptotically anti-de Sitter black holes different boundary conditions at infinity  are discussed in Refs.\ \cite{Moss:2001ga,Siopsis:2007wn}.  We believe that for the UAO and CAO black holes it is convenient to investigate how the QNF change when we impose other boundary conditions at infinity. Another useful extension of our results is to calculate the QNF of CAO black holes whose parameters are not included in the convergence region of the Horowitz Hubeny method plotted in Fig.\ \ref{fig: parameter region}.

\section{Acknowledgments}

This work was supported by CONACYT M\'exico, SNI M\'exico, EDI-IPN, COFAA-IPN, and Research Projects SIP-20110729 and SIP-20111070. R.\ Cordero and I.\ Vega-Acevedo acknowledge financial support from CONACYT research grant no.\ J1-60621-I.

\end{document}